\documentclass[
superscriptaddress,
longbibliography,
tightenlines,
twocolumn,
amsmath,amssymb,
aps,
]{revtex4-1}

\usepackage{amsmath}
\usepackage{amssymb}
\usepackage{bm}
\usepackage{color,graphicx}
\usepackage{colortbl}
\usepackage{slashed}
\usepackage{comment}
\usepackage{mathtools}
\usepackage{booktabs}
\usepackage{xcolor}

\usepackage{url}
\usepackage[section]{placeins}
\usepackage[colorlinks=true,linkcolor=blue,citecolor=blue]{hyperref}

\newcommand{\norm}[1]{\left\lVert#1\right\rVert}

\newcounter{comment}

\usepackage{enumerate}
\usepackage{amssymb}

\begin{document}


\title{Variational autoencoder inverse mapper for extraction of Compton form factors: Benchmarks and conditional learning}

\author{Fayaz Hossen} 
\email{mhoss006@odu.edu}
\affiliation{Department of Computer Science, Old Dominion University, Norfolk, VA 23529, USA.}

\author{Douglas~Adams}  
\email{yax6jr@virginia.edu}
\affiliation{Department of Physics, University of Virginia, Charlottesville, VA 22904, USA.}

\author{Joshua Bautista} 
\affiliation{Department of Physics, University of Virginia, Charlottesville, VA 22904, USA.}

\author{Yaohang Li} 
\email{yaohang@cs.odu.edu}
\affiliation{Department of Computer Science, Old Dominion University, Norfolk, VA 23529, USA.}

\author{Gia-Wei~Chern} 
\email{gchern@virginia.edu}
\affiliation{Department of Physics, University of Virginia, Charlottesville, VA 22904, USA.}

\author{Simonetta Liuti} 
\email{sl4y@virginia.edu}
\affiliation{Department of Physics, University of Virginia, Charlottesville, VA 22904, USA.}


\author{Marie~Bo\"{e}r} 
\affiliation{Department of Physics, Virginia Tech, Blacksburg, VA 24061, USA.}

\author{Marija \v Cui\'c} 
\affiliation{Department of Physics, University of Virginia, Charlottesville, VA 22904, USA}

\author{Michael~Engelhardt} 
\affiliation{Department of Physics, New Mexico State University, Las Cruces, NM 88003, USA}

\author{Gary R. Goldstein} 
\affiliation{Department of Physics and Astronomy, Tufts University, Medford, MA 02155, USA}

\author{Huey-Wen Lin} 
\affiliation{department of physics and astronomy, Michigan State University, East Lansing, MI 48824, USA}

\collaboration{EXCLAIM Collaboration}

\begin{abstract}
Deeply virtual exclusive scattering processes (DVES) serve as precise probes of nucleon quark and gluon distributions in coordinate space. These distributions are derived from generalized parton distributions (GPDs) via Fourier transform relative to proton momentum transfer. QCD factorization theorems enable DVES to be parameterized by Compton form factors (CFFs), which are convolutions of GPDs with perturbatively calculable kernels. Accurate extraction of CFFs from DVCS, benefiting from interference with the Bethe-Heitler (BH) process and a simpler final state structure, is essential for inferring GPDs. This paper focuses on extracting CFFs from DVCS data using a variational autoencoder inverse mapper (VAIM) and its constrained variant (C-VAIM). VAIM is shown to be consistent with Markov Chain Monte Carlo (MCMC) methods in extracting multiple CFF solutions for given kinematics, while C-VAIM effectively captures correlations among CFFs across different kinematic values, providing more constrained solutions. This study represents a crucial first step towards a comprehensive analysis pipeline towards the extraction of GPDs.
\end{abstract}

\maketitle

\allowdisplaybreaks
\newpage
\section{Introduction}
\label{sec:intro}
%
Deeply virtual exclusive scattering processes (DVES) where an electron scatters coherently off a proton target, producing either an additional high energy photon (deeply virtual Compton scattering, DVCS) or a meson (deeply virtual meson production, DVMP), 
are believed to be the most accurate probes of the nucleon quark and gluon distributions in coordinate space.  The latter can be extracted from the non-forward QCD matrix elements between the initial ($p$) and final ($p'$) proton, known as generalized parton distributions (GPDs) \cite{Ji:1996ek,Muller:1994ses,Radyushkin:1997ki}, through Fourier transformation with respect to the proton momentum transfer ($\Delta=p-p'$). 

%
QCD factorization theorems \cite{Collins:1996fb, Collins:1998be, Muller:1994ses} provide a framework where DVES can be parameterized in terms of several Compton form factors (CFFs), which are comprehensively describing all allowed beam-target polarization configurations. Assuming the validity of QCD factorization, CFFs can be written as convolution integrals of GPDs over the longitudinal quark/gluon momentum fraction, $x$, with  kernels/coefficient functions which are calculable in a perturbative QCD collinear approach, similar to the analysis of inclusive deep inelastic scattering. GPDs can therefore be inferred, although only indirectly, from the convolutions given by the measured CFFs (we refer the reader to reviews on the subject in \cite{Diehl:2001pm,Belitsky:2005qn,Kumericki:2016ehc}). 
A quantitative determination of the CFFs from experiment in the asymptotic regime defined by $t/Q^2 << 1$, is, consequently, the first fundamental step in 
the effort to extract hadronic 3D structure from data. The CFFs depend on the kinematic variables, ($t, Q^2, \xi$), where $t=\Delta^2$, $Q^2$, is the electron four-momentum transfer squared defining the QCD scale of the DVES process, and the skewness parameter, $\xi$, measures the fractional longitudinal momentum transfer between the initial and final proton, which is proportional to the kinematic variable, $x_{Bj}$, Bjorken~$x$.

In this paper we focus on the extraction of CFFs from measurements of the DVCS process, which holds a unique role among all DVES reactions because of its two characteristic features: on one side, through the interference with the background Bethe Heitler (BH)
process, where the final state hard photon is radiated from the electron, one can extract the CFFs directly at the amplitude level in linear combinations with known coefficients;   
on the other hand,  because of the presence of only one distinct hadronic blob, DVCS is a cleaner probe, as compared to DVMP, or to similar processes with more than one hadron in the final state, where an additional non leading dependence on the QCD scale, $Q^2$, arises, and needs to be accounted for. 
The DVCS contribution is described by four complex CFFs at leading order: $\mathcal{H}$, $\mathcal{E}$, $\widetilde{\mathcal{H}}$ and $\widetilde{\mathcal{E}}$ (here we will follow the comprehensive formalism given in Refs.~\cite{Kriesten:2019jep,Kriesten:2020wcx}). There are, therefore, eight unknowns given by the real and imaginary parts of the proton CFFs. A sufficient amount of data to grant a sound statistical analysis is available for the case of  scattering from an unpolarized target, while only sparse sets of data exist for other polarization configurations, that could, in principle, provide independent measurements. 
Therefore, the eight unknown real and imaginary components of CFFs need to be extracted from a single polarization observable, thus defining an inverse problem for which there can be an infinite number of solutions. 

The inverse problem of the CFFs extraction from the cross section was the starting point of the analysis in Ref.~\cite{Almaeen:2022ifg} where a variational autoencoder inverse mapper (VAIM)~\cite{9534012} was first introduced to determine CFF solutions including propagated experimental errors, as well as a study of their correlations through the PCA of the latent space. The VAIM analysis of~\cite{Almaeen:2022ifg} presents similarities to the extraction of the parton distribution functions (PDF) parameters from fits to deep inelastic scattering (DIS) data ~\cite{Almaeen:2022ifg}, thus pointing at more general features of this approach. 
%
The main advantage of using an autoencoder for the physics problem at hand is in that through the process of first encoding the input data into a reduced dimensionality subspace, and subsequently decoding them, the autoencoder focuses on and retains the essential, valuable information buried in, and otherwise unattainable from the original data set. 
Through this approach one can provide answers to how much information  is retained through the various layers of the analysis 
experimental data. 
%

To improve our understanding of the working of the VAIM, we carried out a statistical analysis aimed at understanding specifically the extraction of CFFs from DVCS data. Specifically, the inverse problem of CFF extraction is also solved by directly sample solutions using Markov chain Monte Carlo (MCMC) methods. Comparing the results from VAIM and MCMC, we show that the predictions from a VAIM essentially reproduce the distribution used to generate CFF tranining datasets constrained to a given cross section. This analysis thus provides an important benchmark of the VAIM framework (see also discusssion in Ref.\cite{Almaeen2021}). The benchmark of the naive VAIM structure also serves as a reference to highlight the importance of conditional learning in further constraining the CFF predictions. By including kinematic variables as additional control parameters to both forward and backward mappers, nontrivial constraints on the inverse-problem solutions are introduced through a global optimization mechaism. We demonstrate that highly structured distributions are obtained for some CFFs from the predictions of conditional-VAIM (C-VAIM).



This paper is organized as follows: in Section ~\ref{sec:background} we give a theoretical description of the specific inverse problem for DVCS; in Section \ref{sec:VAIMand MCMC} we describe the basic elements of the VAIM algorithm, specifically in Section  ~\ref{sec:vaim-basics} where we motivate the VAIM architecture used to solve the inverse problem of CFF extraction, and in Section \ref{sec:MCMC} where we explore the connection to MCMC; Section~\ref{sec:C-VAIM} discusses the conditional VAIM (C-VAIM). We draw our conclusions in Section ~\ref{sec:conclusions}.

\bigskip


\section{CFF extraction from DVCS as an inverse problem}
\label{sec:background}

\begin{figure*}[t]
    \centering
    \includegraphics[width=1.35\columnwidth]{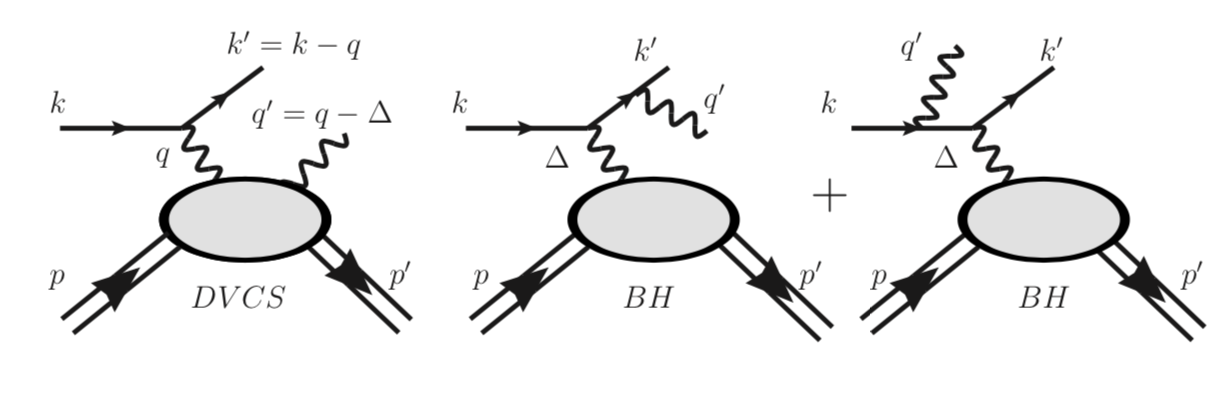}
    \caption{Feynman diagrams describing the amplitudes for DVCS  (left) and BH (right). Both contribute to the  photon electro-production $ep \to ep\gamma$.}
    \label{fig:Feyn}
\end{figure*}

In the QCD factorization framework \cite{Collins:1996fb, Collins:1998be, Muller:1994ses} which is assumed to be valid at sufficiently high energies, and assuming the smallness of the ratio between the four-momentum difference squared, $t=(p-p')^2= \Delta^2$, and the QCD scale of the process defined by the lepton ofur-momentum transfer, $Q^2=-(k-k')^2= -q^2$, the DVCS amplitude (Figure \ref{fig:Feyn}) can be written as a linear combination of the four leading twist CFFs,  $\mathcal{H}$, $\mathcal{E}$, $\widetilde{\mathcal{H}}$ and $\widetilde{\mathcal{E}}$ ~\cite{Kriesten:2019jep,Kriesten:2020wcx}, with exactly calculable kinematic coefficients.


Aside from DVCS, the total cross section for photon electro-production also includes the background BH process which gets summed coherently to form the measured cross section as,
$\sigma_{\rm TOT} \propto \left| \mathcal{T}_{  BH} + \mathcal{T}_{  DVCS} \right|^2$, leading to,
\begin{eqnarray}
	\sigma_{\rm TOT} = \sigma_{  BH} + \sigma_{  DVCS} + \sigma_{\mathcal{I}}, 
\end{eqnarray}
For the unpolarized cross section, explicit expressions for the three cross section components at leading twist are given by \cite{Kriesten:2019jep,Kriesten:2020wcx},
\begin{widetext}
\begin{subequations}
\label{eq:sigmatot_eqs}
\begin{eqnarray}
\label{eq:BHxsec}
\sigma_{BH}[KIN] &=& \frac{\Gamma}{t}\Big[A_{BH}[KIN]\big(F_1(t)^2 + \tau F_2(t)^2 \big) + B_{BH}[KIN] \tau (F_1(t)+F_2(t))^2\Big]\\
\label{eq:DVCSxsec}
\sigma_{DVCS}[KIN]  & = & \frac{\Gamma}{Q^2(1-\epsilon)} \nonumber \\
& \times & \Bigg\{ 2( 1 - \xi^2) \bigg( \left[\Re e \mathcal{H}(x_{Bj},t,Q^2)\right]^2 + \left[\Im m \mathcal{H}(x_{Bj},t,Q^2)\right]^2 + \big[\Re e \widetilde{\mathcal{H}}(x_{Bj},t,Q^2) \big]^2 + \big[\Im m \widetilde{\mathcal{H}}(x_{Bj},t,Q^2)\big]^2\bigg) \nonumber \\
& + & \frac{t_0-t}{2 M^2}\bigg( \left[\Re e \mathcal{E}(x_{Bj},t,Q^2) \right]^2 + \left[\Im m \mathcal{E}(x_{Bj},t,Q^2)\right]^2 + 
\big[ \xi \, \Re e  \widetilde{\mathcal{E}}(x_{Bj},t,Q^2)\big]^2 +  \big[\xi  \,\Im m  \widetilde{\mathcal{E}}(x_{Bj},t,Q^2) \big]^2 \bigg)   \nonumber \\
&  - & 4 \xi^2\bigg(\Re e \mathcal{H}(x_{Bj},t,Q^2) \, \Re e \mathcal{E}(x_{Bj},t,Q^2) + \Im m \mathcal{H}(x_{Bj},t,Q^2) \, \Im m \mathcal{E}(x_{Bj},t,Q^2) \nonumber \\
  & \quad\quad+ &  \Re e \widetilde{\mathcal{H}}(x_{Bj},t,Q^2) \, \Re e \widetilde{\mathcal{E}}(x_{Bj},t,Q^2) 
+ \Im m \widetilde{\mathcal{H}}(x_{Bj},t,Q^2) \, \Im m \widetilde{\mathcal{E}}(x_{Bj},t,Q^2) \bigg)  \Bigg\} ,
 \\
\label{eq:Ixsec}
\sigma_{\cal I} [KIN] &=& \frac{\Gamma}{Q^{2}} \Big[ A_{\mathcal{I}}[KIN]\Big(F_{1}(t) \, { \Re e \mathcal{H}}(x_{Bj},t,Q^2) + \tau F_{2}(t) \, { \Re e \, \mathcal{E}}(x_{Bj},t,Q^2) \Big) \nonumber \\
&+& B_{\mathcal{I}}[KIN] (F_1(t)+F_2(t))  \, \Big( { \Re e \mathcal{H}}(x_{Bj},t,Q^2)  + { \Re e \mathcal{E}}(x_{Bj},t,Q^2) \Big) \nonumber \\
& + & C_{\mathcal{I}}[KIN](F_1(t)+F_2(t)) \, { \Re e \widetilde{\mathcal{H}}}(x_{Bj},t,Q^2) \: \Big] \, .
\end{eqnarray}
\end{subequations}
\end{widetext}
where all three components of  the cross section are written in terms the kinematic variables $[KIN]= (s, Q^2, x_{Bj}, t, \phi)$, defined as,
\footnote{At the leading order considered here, the DVCS contribution in Eq.\eqref{eq:DVCSxsec} technically depends on the subset $(s, Q^2, x_{Bj}, t)$, that is it is $\phi$ independent (see \cite{Kriesten:2020wcx} for a detailed discussion of this point).}
\begin{itemize}
    \item $s=(k+p)^2$ is the electron-proton center of mass energy squared. In a fixed target configuration $s=\sqrt{2 M E}$, where $E$ is the energy of the initial electron, and $M$ is the nucleon mass.
    \item $Q^2 = -(k-k')^2 = - q^2$ is the four-momentum transfer squared between the incoming and outgoing electrons,
    \item $x_{Bj} = Q^2 / 2(pq)$ is Bjorken-$x$. In the asymptotic limit, disregarding $t/Q^2$ ans $M^2/Q^2$ corrections, $x_{Bj}$ is written in terms the skewness parameter, $\xi = - (\Delta q)/[(pq)+(p'q)]= x_{Bj}/(2- x_{Bj})$
    \item $t = (p- p')^2 = (q'-q)^2$ is the four-momentum transfer squared between the initial and final protons ($q'$ is the final photon four-momentum),
    \item $\phi$ is the azimuthal angle between the planes defined by the electron momenta, by the final proton, $p'$, and photon, $q'$.
\end{itemize}
In this paper we use unpolarized electrons scattering off an unpolarized proton data, therefore the cross section does not depend on an additional azimuthal angle, $\phi_S$,  stemming from the proton spin orientation. 
The variables $\Gamma$, $\epsilon$ (the ratio of virtual photon, $q$, longitudinal over transverse polarization), and $t_0$ (the minimum kinematically allowed value of $t$, corresponding to $\Delta_T=0$), respectively, read, 
\begin{equation}\Gamma = \frac{\alpha^3 (2 \pi) }{16\pi^2 (s-M^2)^2 \sqrt{1+\gamma^2}\, x_{Bj} },
\end{equation}
where $\alpha$ is the electromagnetic fine structure constant, and $\gamma^2 = 4M^2x_{Bj}^2/Q^2$,
and,
\begin{equation}
\epsilon = \frac{1-y-\frac{1}{4} y^2 \gamma^2}{1-y+\frac{1}{2} y^2+\frac{1}{4} y^2 \gamma^2},
\end{equation}
with $y=(pq)/(pk)= Q^2/(x_{Bj} s)$,
\begin{equation}
    t_0 = - \frac{4 M^2 \xi^2}{(1-\xi)}  .
\end{equation}
$F_1$ and $F_2$ are the Dirac and Pauli proton elastic form factors which we considered known to high accuracy in the $t$ regime accessed in this paper. 
The coefficients $A_{BH}$, $B_{BH}$, as well as $A_{\cal I}$, $B_{\cal I}$  $C_{\cal I}$, have rather lenghtly expressions that we do not report here, but refer the reader to Ref.\cite{Kriesten:2020wcx} where they are explicitely written and discussed. 
The CFFs stem from QCD factorization treated at leading order in this paper, and they enter the DVCS amplitude and consequently the expressions for both the interference and the DVCS terms.
%
Details of the formalism adopted in this paper and a comparison with other work {\it e.g.} in \cite{Belitsky:2001ns, Belitsky:2010jw,Braun:2012hq} can be found in \cite{Kriesten:2019jep, Kriesten:2020apm,Kriesten:2020wcx}. An alternative formalism was also presented in \cite{Guo:2021gru}.  

Given a set of kinematic parameters, extracting the eight CFF parameters, {\it i.e.} the  real and imaginary parts of $\mathcal{H}$, $\mathcal{E}$, $\widetilde{\mathcal{H}}$, and $\widetilde{\mathcal{E}}$, constitutes an {\it inverse problem}. 

In our analysis, we arrange the CFFs for any given kinematics, $[KIN]$, into a dimension, $D=8$, components vector, namely, 
\begin{equation}
    \begin{aligned}
        \bm{x}= \vec{x} = \begin{bmatrix}
        x_1 \\
        x_2 \\
        x_3 \\
        x_4  \\
        x_5 \\
        x_6 \\
        x_7 \\
        x_8
        \end{bmatrix}
= 
\begin{bmatrix}
        \Re e\,\mathcal{H} \\
        \Im m\, \mathcal{H} \\
        \Re e\,\mathcal{E} \\
        \Im m\, \mathcal{E} \\
        \Re e\,\widetilde{\mathcal{H}} \\
        \Im m\, \widetilde{\mathcal{H}} \\
        \Re e\,\widetilde{\mathcal{E}} \\
        \Im m\, \widetilde{\mathcal{E}} 
                \end{bmatrix}
    \end{aligned}
\end{equation}
Next, we define a function ${\cal F}(\bm{x})$ of the CFFs as the sum of the two cross section components containing CFFs in Eqs.~(\ref{eq:Ixsec}) and (\ref{eq:DVCSxsec}), respectively. 
The extraction of CFF here is defined as solving the following equation for $\bm x$:
\begin{eqnarray}
	\label{eq:inverse-prob}
	\sigma_{\rm exp} = \mathcal{F}(\bm{x}).
\end{eqnarray}
Here $\sigma_{\rm exp}$ is evaluated as,
\begin{equation}
    \sigma_{\rm exp} = \sigma_{TOT}^{\rm exp} - \sigma_{BH} 
\end{equation}
where $\sigma_{TOT}^{\rm exp}$, is the experimentally measured cross section; the background BH contribution, $\sigma_{BH}$, is subtracted by calculating this terms using Eq.\eqref{eq:BHxsec}, with the parametrization of the nucleon form factors from \cite{Kelly:2004hm} 


From Eqs.\eqref{eq:sigmatot_eqs} one cans see that ${\cal F}$ is given by the following quadratic function 
\begin{equation}
    {\cal F}(\bm{x}) = \bm a \cdot \bm{x} + \bm{x}^{\intercal} \cdot \mathbb{M} \cdot \bm{x},
\end{equation} 
where the elements of ${\bm a}$ and $\mathbb{M}$ are given by linear combinations of the various kinematics coefficients from Eqs.\eqref{eq:sigmatot_eqs}.
Yet, even with this relatively simple form, there can be an infinite number of solutions $\bm{x}$ to Eq.~(\ref{eq:inverse-prob}). 

In previous work~\cite{VAIM-CFF24}, a deep-learning approach based on the variational autoencoder inverse mapper (VAIM) was utilized to solve the inverse problem of CFF extraction. 
The present paper is dedicated to studying the VAIM approach, and its benchmark against solutions of the inverse problem based on Monte Carlo sampling.

\section{CFF extraction based on VAIM and MCMC samplings}
\label{sec:VAIMand MCMC}

\subsection{VAIM basics}
\label{sec:vaim-basics}

ML and data science methods have revolutionized scientific research in a diverse range of fields. The universal approximation theorem~\cite{Cybenko89,Hornik89} provides the theoretical basis for utilizing deep-learning neural networks to approximate complex high-dimensional functions. In particular, with easy accesses to large amount of experimental and numerical data, accurate learning models can be obtained by combining the universal approximation capabilities of neural net with the data-driven approach for their training. The rapid progress of ML methods is further galvanized by the development of robust and efficient optimization algorithms and graphic processing units (GPU).

\begin{figure}[]
\includegraphics[width=0.99\columnwidth]{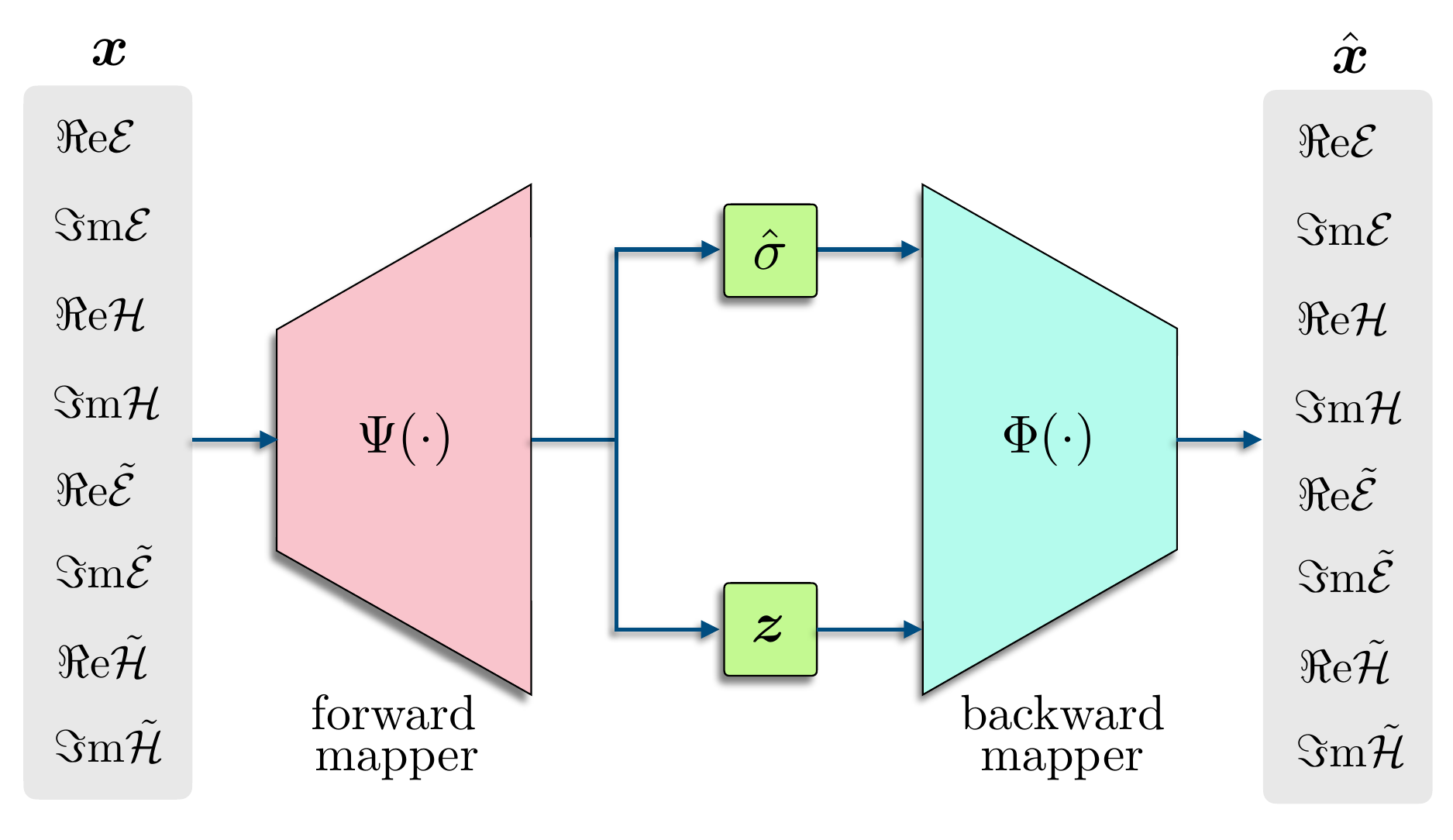}
\caption{\label{fig:vaim-schematic} Structure of a VAIM for solving the inverse problem of CFF extraction. The ML model consists of two neural networks: the forward mapper $\Psi(\cdot)$ and the inverse mapper~$\Phi(\cdot)$. The latent variables $\bm z$ are introduced to capture the lost information in the many-to-one function $\mathcal{F}(\bm x)$ in the inverse problem. }
\end{figure}

Here we first review a novel deep-learning approach, known as the Variational Autoencoder Inverse Mapper (VAIM)~\cite{VAIM} for solving the inverse problem of CFF extraction. VAIM is a general end-to-end neural network architecture designed for addressing inverse problems using an autoencoder-based approach~\cite{autoencoder}. The basic structure of a VAIM is shown in FIG.~\ref{fig:vaim-schematic}; it comprises two key deep neural networks: an encoder and a decoder. The encoder approximates the forward mapping, while the decoder approximates the backward mapping.

In VAIM, the inverse problem is treated as a statistical inverse problem, with parameters modeled as random variables. Rather than producing deterministic estimates for given observables, VAIM approximates the probability distribution of the parameters. A critical component of VAIM is the variational latent layer, positioned between the encoder and the decoder. This layer is designed to learn the patterns of the parameter distributions and serves as part of the encoder's output and the decoder's input.

Unlike the hidden layers in the encoder or decoder, the latent layer is constrained to follow specific well-known distributions, such as Gaussian or uniform distributions, which is calculated from the Kullback-Leibler (KL) divergence between the two distributions, to facilitate variational inference~\cite{VariationalInference}.   Solving a statistical inverse problem involves sampling from the latent distribution for a given observable to obtain the corresponding parameter distributions. Parameters with high sensitivity will yield narrow distributions, while those with low sensitivity will produce wider distributions. The overall loss function of consists of three terms: (i) a reconstruction loss which measures how well the reconstructed observables match the original observables in the forward mapper, (ii) the error between the predicted parameters and the true parameters, and (iii) a regularization term to ensure that the latent space to follow the prior distribution, which is calculated from the Kullback-Leibler (KL) divergence between the two.

While the basic structure of a VAIM is similar to that of a variational auto-encoder (VAE)~\cite{2019arXiv190602691K,vae}, instead of merely encoding the input data into latent variables, the first neural net, called a forward mapper $\Psi(\cdot)$, also provide an approximation to the function $\mathcal{F}(\bm x)$ to be learned. In our case, the forward mapper is defined as 
\begin{eqnarray}
	\Psi: \bm x \to ( \hat{\sigma}, \bm z), 
\end{eqnarray}
where $\hat{\sigma}$ is the target cross section related to the CFF through the function $\mathcal{F}(\bm x)$ in Eq.~(\ref{eq:inverse-prob}) and $\bm z = (z_1, z_2, \cdots)$ is a set of latent variables. The introduction of the latent variables can be understood as follows. As $\mathcal{F}(\bm x)$ is a nonlinear function, different input CFFs, say $\bm x_A$ and $\bm x_B$, would result in the same cross section. The latent variables are then introduced to capture the lost information in this many-to-one mapping, i.e. the forward mapper is designed to produce distinct latent variables $\bm z_A$ and $\bm z_B$ for the two different inputs. 

With the additional information about different input CFFs stored in the latent space, the inverse mapper $\Phi(\cdot)$, which is similar to the decoder in a VAE, is trained to reconstruct the CFFs:
\begin{eqnarray}
	\Phi: (\sigma, \bm z) \to \hat{\bm x}.
\end{eqnarray}
Here the symbol $\hat{\bm x}$ is used to denote an ML prediction for the intended variables. 

The training of VAIM is also similar to that of conventional VAE. Formally, the forward mapper is designed to approximate the posterior distribution $p(\bm z \, | \, \bm x, \sigma)$, while the backward mapper learns to approximate the likelihood distribution $p(\bm x, \sigma \, | \, \bm z)$. Since the true posterior is intractable, the variational inference is employed to approximate the true distribution $p(\bm z \, | \, \bm x, \sigma)$ by another tractable distribution $q(\bm z\, | \, \bm x, \sigma)$ implemented in the forward mapper. To this end, the KL divergence between these two distributions is introduced as a measure of their difference.

Employing the variational inference principle developed for the theory of VAE, the training of a VAIM amounts to the minimization of the following loss function
\begin{eqnarray}
	\label{eq:loss_func}
	L = \norm{\sigma - \hat{\sigma}}_2^2 + \norm{\bm x - \hat{\bm x}}_2^2 + {\rm KL}\bigl(q(\bm z \, | \, \bm x, \sigma) \, || \, p(\bm z) \bigr),
\end{eqnarray}
where $\norm{\cdot}_2$ is the L$_2$ norm and ${\rm KL}(q, p)$ denotes the KL divergence between two distributions $q$ and $p$. The first two terms in the above loss function represent the forward mapping error and the reconstruction error, respectively, and $p(\bm z)$ is a true prior distribution, which is often chosen to be a tractable, easy-to-generate distribution such as a normal distribution. Once a VAIM is successfully trained, the inverse mapper can then be used as a generative model to produce CFFs by sampling the latent variables $\bm z$ with respect to a given cross section $\sigma$. 

The distribution of the extracted solutions $\hat{\bm x}$ for a given cross section depends crucially on the training dataset. Indeed, the training dataset in the form of $\left\{ \bm x^{(m)}, \sigma^{(m)} \right\}$, where $m = 1, 2, \cdots, N_{\rm data}$, are in general obtained by sampling a distribution function $\Pi(\bm x)$ in the 8-dimensional CFF space.  In our case, each of the CFFs is sampled uniformly from a predefined domain~$[l_j, u_j]$, which means that $\Pi(\bm x) = \Pi_0   = $ const. for $\bm x$ within the hyper-cuboid defined by these intervals and $\Pi(\bm x) = 0$ otherwise. 

More relevant to the VAIM prediction is the distribution function $\pi(\bm x; \sigma)$ of CFFs constrained to a manifold defined by Eq.~(\ref{eq:inverse-prob}). This distribution function naturally depends on the underlying $\Pi(\bm x)$ used to sample the CFFs. Intuitively, the forward mapper can be viewed as a transformation of ``random variables" that convert a distribution function $\pi(\bm x;  \sigma)$ into a prior distribution $p(\bm z)$ which is selected to be a normal distribution in our implementation. Similarly, for a given cross section $\sigma$, the inverse mapper converts the distribution $p(\bm z)$ back to $\pi(\bm x;  \sigma)$ in the CFF space.  This also means that the extracted CFFs $\hat{\bm x}$ from the inverse mapper for a given~$\sigma$ are expected to follow the distribution of the constrained~CFFs, i.e. $\hat{\bm x} \sim \pi(\bm x; \, \sigma)$.

\begin{figure}[]
\includegraphics[width=0.95\columnwidth]{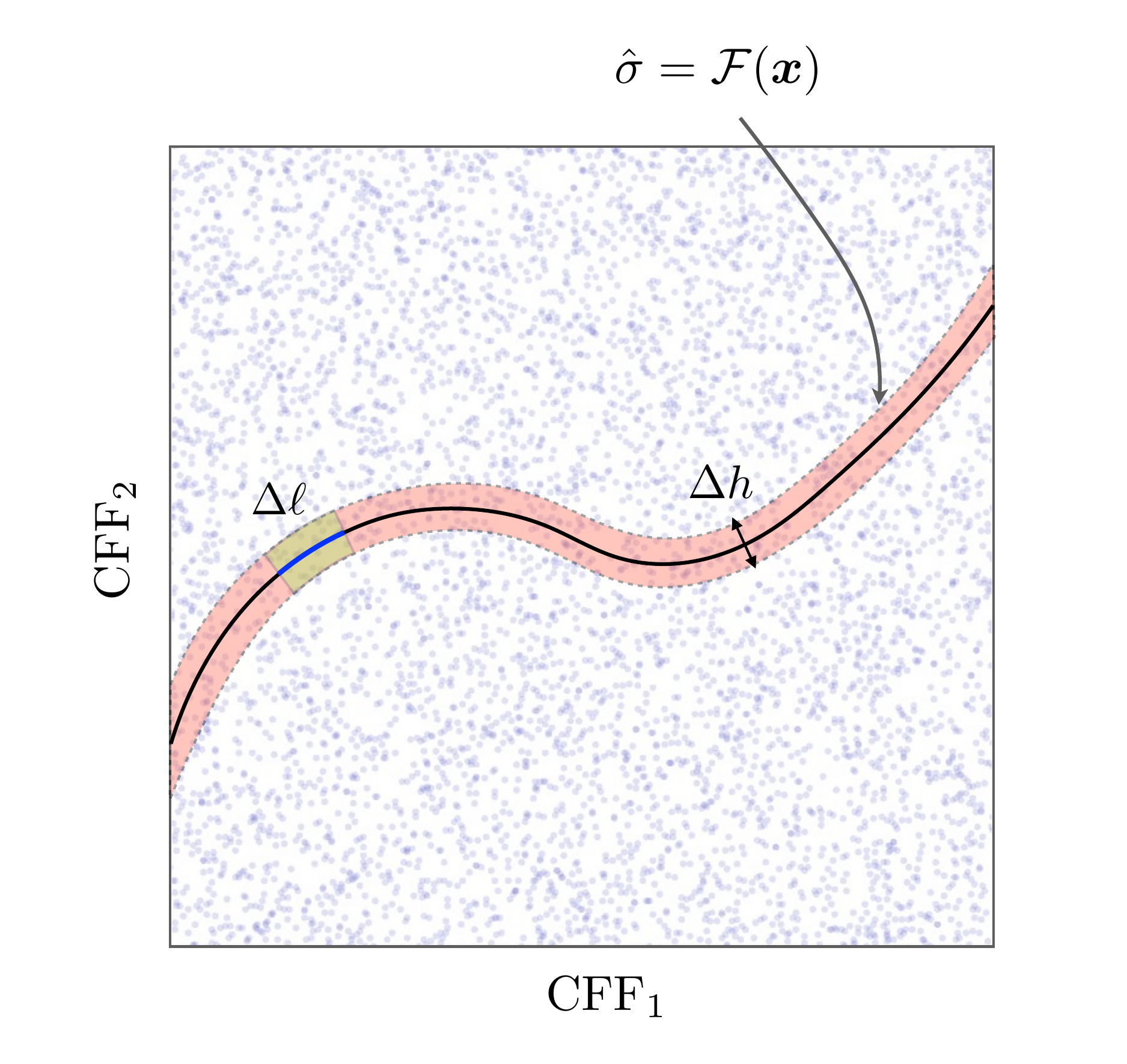}
\caption{\label{fig:sampling-surface} Illustration of uniform distribution $\pi(\bm x; \sigma)$ on the constraint manifold, which corresponds to the solid curve, from a uniform sampling of CFFs in a square.}
\end{figure}

\subsection{MCMC sampling of the constraint manifold}
\label{sec:MCMC}

To benchmark the CFF predictions of VAIM, one way is to perform direct Monte Carlo samplings of the distribution function $\pi(\bm x; \sigma)$. To relate this distribution to the density function $\Pi(\bm x)$ used to sample the CFFs, we first note that the manifold defined by the constraint Eq.~(\ref{eq:inverse-prob}) corresponds to a $(D-1)$-dim ``surface" embedded in a $D$-dim Euclidean space of CFFs. Assuming that a total of $\mathcal{N}_0$ CFFs are sampled using the distribution $\Pi(\bm x)$, the number density of sampled points in the $D$-dim space is simply $\rho(\bm x) = \mathcal{N}_0 \Pi(\bm x)$. Consider a  $D$-dim hyper-cylindrical volume element $d^D V = \Delta h \times d^{D-1} A$ centered at $\bm x$ on the constraint surface. The base of the cylinder corresponds to an ``area'' element $d^{D-1}A$ on the surface, and its height $\Delta h$ is along the direction of the surface normal $\hat{\bm n} \propto \nabla \mathcal{F} $. The number of sample points within this volume element is $N = \rho(\bm x) d^D V$, and the ``surface" density of sample points is $\rho_s(\bm x) = N / d^{D-1}A = \rho(\bm x) \Delta h$. The distribution $\pi(\bm x; \sigma)$ is expected to be proportional to the surface density. Using the normalization condition to determine the proportional coefficient, we have
\begin{eqnarray}
	\pi(\bm x; \sigma) = \lim_{\Delta h \to 0} \frac{\rho_s(\bm x)}{\int_{\Sigma} \rho_s(\bm x') d^{D-1} A} =  \mbox{const.} \times  \Pi(\bm x), \quad
\end{eqnarray}
where the integral in the denominator is over the constraint manifold. The ``surface" distribution is simply the bulk distribution restricted to the constraint Eq.~(\ref{eq:inverse-prob}).

For the special case of uniform sampling, we have $\Pi(\bm x) = \Pi_0$, a constant, for CFFs within the predefined hypercuboid. The above equation indicates the CFFs that produce the same cross section are also uniformly distributed within the constraint manifold. To understand this result intuitively, we consider the simplified hypothetical situation of $D = 2$ as shown in FIG.~\ref{fig:sampling-surface}. The two CFFs are assumed to be uniformly sampled within the square. The solid curve corresponds to the 1D manifold defined by the constraint Eq.~(\ref{eq:inverse-prob}). Consider adding an infinitesimal width $\Delta h$ to the curve, making it a strip. Since the CFFs are uniformly sampled in the 2D background, their distribution within the strip is also uniform. For example, the number of sample points is proportional to the length $\Delta \ell$ of a line segment. By letting the thickness $\Delta h \to 0$, this then translates to a uniform distribution on the curve. 

Given the distribution function $\pi(\bm x; \sigma)$ for the constraint manifold, the Metropolis-Hastings importance sampling algorithm can be used to sample CFFs on the ``surface". However, a direct sampling of CFFs, which requires an efficient local updates of sampling points $\bm x \to \bm x'$ on the surface, is infeasible because of the nontrivial constraint imposed by Eq.~(\ref{eq:inverse-prob}). A practicable approach is to introduce a parametrization or local coordinates $\bm \xi = (\xi_1, \xi_2, \cdots, \xi_{D-1})$ for the $(D-1)$-dim surface, i.e. $\bm x = \bm x(\xi_1, \xi_2, \cdots, \xi_{D-1})$. With this representation any random walk algorithm in the parametrization domain can be used for the local updates.

To apply the Metropolis Hastings method, one needs to express the surface distribution $\pi(\bm x; \sigma)$ in terms of the independent local coordinates~\cite{Liu22}. To this end, we note that probability conservation implies $\pi(\bm x; \sigma) d^{D-1} A = \tilde{\pi}(\bm \xi) d\xi_1 d\xi_2 \cdots d\xi_{D-1}$, where $\tilde{\pi}(\bm \xi)$ denotes the effective density function in the parametrization domain, and the volume elements in the $\bm x$ and $\bm \xi$ spaces are related by
\[
	d^{D-1} A = \sqrt{|{\rm det}(\mathtt{g})|} d\xi_1 d\xi_2 \cdots d\xi_{D-1},
\]
where $\mathtt{g}$ is the metric tensor defined as $g_{mn} = \frac{\partial \bm x}{\partial \xi_m} \cdot \frac{\partial \bm x}{\partial \xi_n}$. The effective distribution function in local coordinates is then related to the one in CFF space by a scaling factor dependent on the metric tensor of the surface:
\begin{eqnarray}
	\tilde{\pi}(\bm \xi) = \sqrt{|{\rm det}(\mathtt{g}) | } \, \pi(\bm x; \sigma).
\end{eqnarray}
Importantly, for the special case of uniform sampling, the effective distribution is simply proportional to the geometrical scaling factor: $\tilde{\pi}(\bm \xi) = {\rm const.} \times \sqrt{ |{\rm det}(\mathtt{g})|}$, where the constant is determined by the normalization condition.  A Markov-Chain Monte Carlo (MCMC) simulation can be carried out using the following transition probability based on the Metropolis-Hastings algorithm,
\begin{eqnarray}
	P(\bm \xi \to \bm \xi') = \pi_{\rm step}(|\bm \xi' - \bm \xi|) \, {\rm min}\!\left(1, \sqrt{ \left| \frac{ {\rm det}\left[ \mathtt{g}(\bm \xi') \right] }{ {\rm det}\left[\mathtt{g}(\bm \xi) \right] } \right| } \Theta(\bm \xi') \right). \nonumber \\
\end{eqnarray}
Here the $\pi_{\rm step}(|\bm \ell|)$ is a probability distribution function used to determine the step size of random walk in the local coordinates, i.e. $\bm \xi' = \bm \xi + \bm \ell$, and the function $\Theta(\bm \xi')$ is to ensure that the walker is within the prior hyper-cuboid, i.e. $\Theta(\bm \xi') = 1$ if $\bm \xi'$ is within the redefined domain, and zero otherwise.

\subsection{Benchmark of VAIM }

\label{sec:benchmark-vaim}

\begin{figure*}[]
\includegraphics[width=1.99\columnwidth]{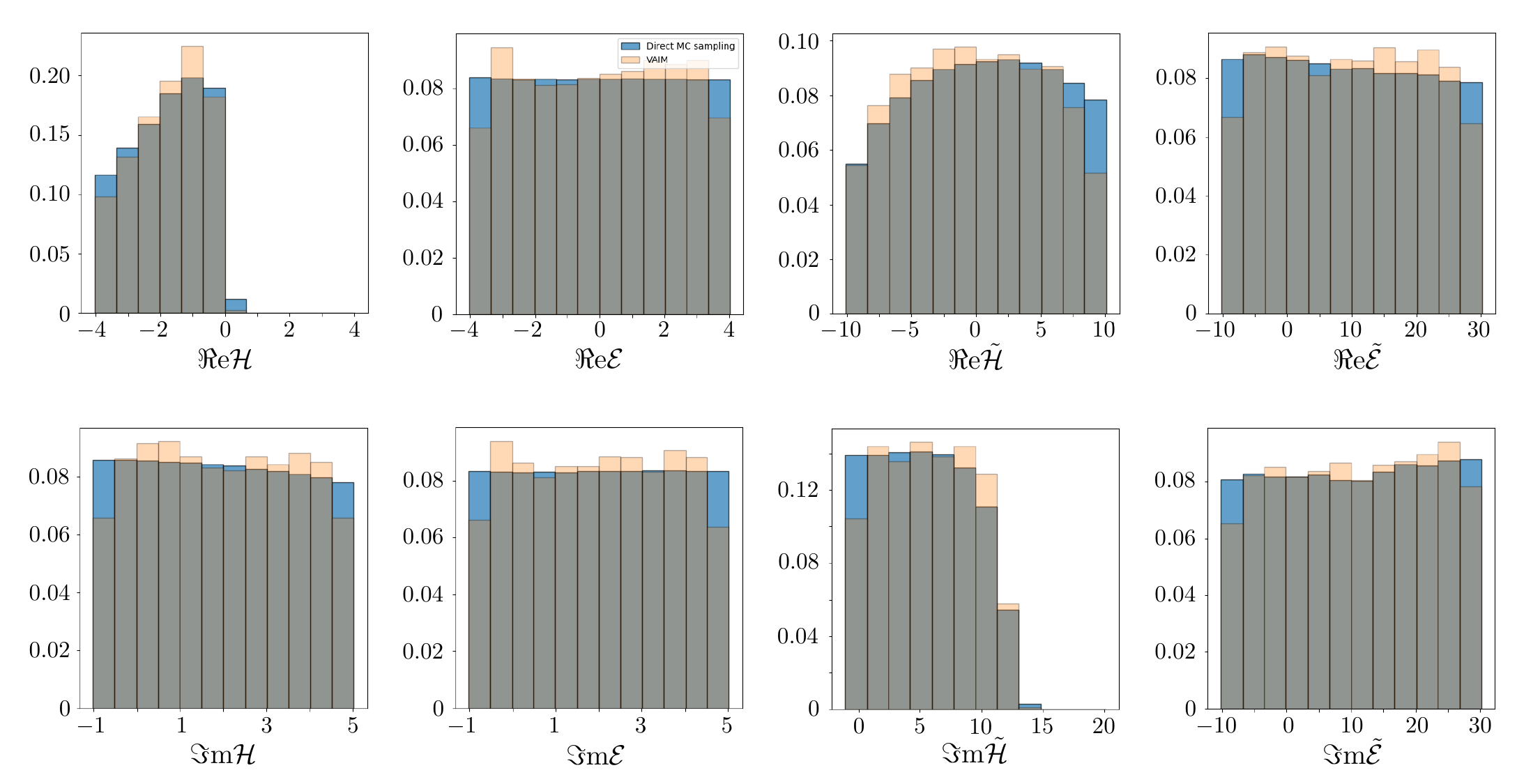}
\caption{\label{fig:hist1} Histogram of CFFs from sampling based on VAIM versus the MCMC method. All results are shown for a specific kinematics for the unpolarized cross section at $Q^2 = 1.82$~GeV$^2$, $t = -0.172$~GeV$^2$, $x_{Bj} = 0.343$, and $E_b = 5.75$~GeV.}
\end{figure*}

\begin{table}
\setlength\extrarowheight{5pt}
\begin{tabular}{| c | c | c | c |}
	\hline
	$\Re e \mathcal{H} \!:\!  [-4, 4]$ & $\Re e \mathcal{E}\!:\!  [-4, 4]$ & $\Re e \widetilde{\mathcal{H}}\!:\!  [-10, 10]$ & $\Re e \widetilde{\mathcal{E}}\!:\!  [-10, 30]$ \\ 
	\hline
	$\Im m \mathcal{H}\!:\! [-1,5]$ & $\Im m \mathcal{E}\!:\!  [-1, 5] $ & $\Im m \widetilde{\mathcal{H}}\!:\!  [-1, 20] $ & $\Im m \widetilde{\mathcal{E}}\!:\! [-10, 30]$ \\
	\hline
\end{tabular}
\caption{\label{tab:prior-regions} Region priors for sampling CFFs and the training of VAIM.}
\end{table}

Here we apply both VAIM and MCMC methods to extract CFFs at the kinematics parameters $Q^2 = 1.82$~GeV$^2$, $t = -0.172$~GeV$^2$, $x_{Bj} = 0.343$, and $E_b = 5.75$~GeV. With sufficient number of sampling, the MCMC method provides an exact approach to the inverse problem, hence serving as a benchmark for the VAIM predictions. To this end, we first train a VAIM based on the selected kinematics. The dataset is obtained by first uniformly sampling CFFs from prior regions summarized in Table~\ref{tab:prior-regions}. For each sampled CFFs $\bm x^{(m)}$, Eqs.~(\ref{eq:Ixsec}) and~(\ref{eq:DVCSxsec}) are used to compute the corresponding cross section $\sigma^{(m)}$ based on the chosen kinematics parameters. The dataset consisting of pairs $\left\{ \bm x^{(m)}, \sigma^{(m)} \right\}$ are used to train a VAIM based on the loss function defined in Eq.~(\ref{eq:loss_func}). Once the VAIM is successfully trained, the inverse mapper $\Phi(\cdot)$ is used to sample CFFs for a given cross section value. Specifically, for each prediction, a set of latent variables $\bm z$ are sampled from the prior distribution $p(\bm z)$, which is chosen to be a normal distribution. The predicted CFFs are obtained by sending both the cross section and latent variable as input to the inverse-mapper neural network.

In our implementation of the VAIM, both forward and backward mappers consist of five fully connected layers, each with 1024 hidden nodes. The latent layer, a critical hyperparameter, has 200 dimensions, ensuring it retains all necessary information. VAIM is trained using the Adam optimizer with a learning rate of \( 1 \times 10^{-5} \), and training continues until the reconstruction error falls below \( 10^{-3} \). This hybrid approach integrates supervised learning for parameter-observable pairs with unsupervised learning for the posterior distribution, enabling VAIM to generalize across diverse inverse problems and effectively manage cases with non-unique solutions.

For the parametrization of the constraint manifold in MCMC simulations, we use the first 7 CFF as the local coordinates to parametrize the constraint manifold, i.e. $x_m = \xi_m$ for $m = 1, 2, \cdots, 7$, while the last CFF $x_8$ is obtained by solving a quadratic equation obtained by substituting the first seven CFFs into Eq.~(\ref{eq:inverse-prob}). The random step $\bm \ell$ is uniformly sampled from within a $(D-1)$-ball of radius $r_0 = 0.1$. The proposed update is rejected if the walker's ``position" $\bm x'$ is outside the prior hyper-cuboid. The combination of this hard constraint and the importance sampling criterion (to account for uniform sampling on the surface) gives an acceptance rate of roughly 55\%. The overall results, however, do not depend critically on the step-size. For a typical MCMC simulations, with 10000 random steps in between two sampled CFFs, a total of $10^6$ CFFs can be efficiently sampled.

\begin{figure}[]
\includegraphics[width=0.99\columnwidth]{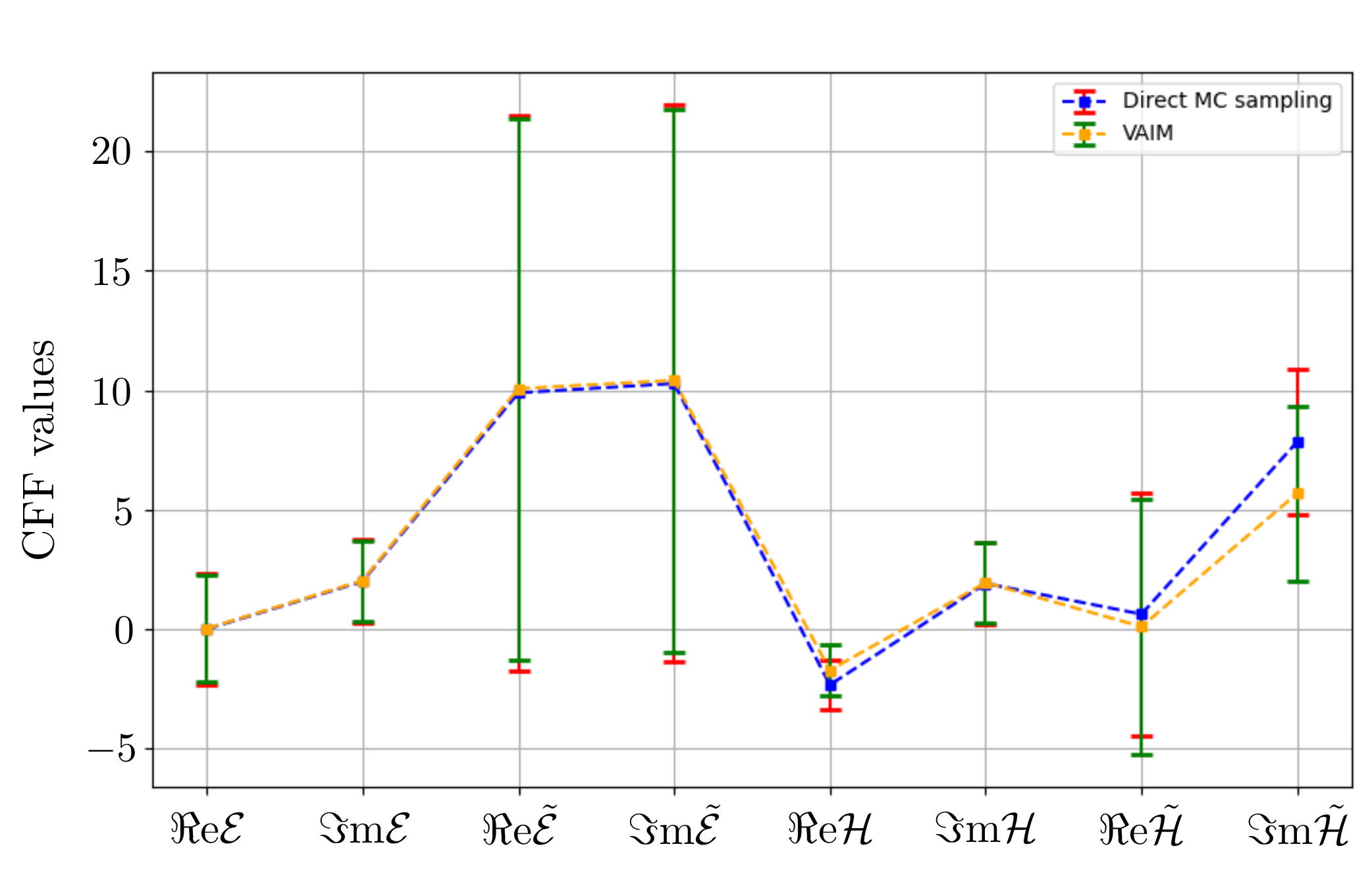}
\caption{\label{fig:cmp1} Comparison of CFF central values and standard deviations extracted from VAIM and MCMC sampling. The error bars indicate one standard deviation obtained from the corresponding histogram data in FIG.~\ref{fig:hist1}.  }
\end{figure}

The histograms of the 8 CFFs sampled using the VAIM and MCMC are both shown in FIG.~\ref{fig:hist1} for comparison. The predictions from VAIM agree very well with those sampled by the numerically exact MCMC method. An excellent overall agreement is also illustrated in FIG.~\ref{fig:cmp1} which shows the central value and the one standard deviation of the 8 CFFs. In addition to providing an excellent benchmark of the VAIM approach. Our comparative study here also clarifies the nature of VAIM predictions: the output $\hat{\bm x}$ of the inverse mapper obtained by sampling the Gaussian-distributed latent space corresponds to a uniform distribution on the constraint surface. 

On the other hand, it is worth noting that most of the sampled CFFs, except for $\Re e \mathcal{H}$, $\Re e \widetilde{\mathcal{H}}$ and $\Im m \widetilde{\mathcal{H}}$, exhibit a nearly uniform distribution within the prior region used for the sampling of CFF training datasets. The fact that most of the predicted CFFs are all over the prior domains highlights more of the ill-defined nature of the inverse problem, instead of a lack of predictive power of the VAIM or MCMC methods. However, some nontrivial features are still obtained for the three CFFs $\Re e \mathcal{H}$, $\Re e \widetilde{\mathcal{H}}$ and $\Im m \widetilde{\mathcal{H}}$. For example, the predicted values of $\Re e \mathcal{H}$ are predominantly negative even though solutions of positive $\Re e \mathcal{H}$ are possible. These results points to nontrivial kinematic constraints in the DVCS cross section formulas. 

Formally, the central values of the VAIM predicted $\hat{\bm x}$ are designated as the extracted CFFs.  However, as discussed above, the large uncertainties in the predictions indicates that the CFF extraction from a single unpolarized cross section is highly under-constrained inverse problem. Moreover, the lack of any structures in most histogram also points to the fact that the predicted results strongly depend on the priors used for generating the CFF data. The extracted values of CFFs are then simply artifacts of the prior domains used, instead of resulting from the underlying physics.   In the next section, we show how the uncertainty of CFF extraction can be significantly reduced using the technique of conditional learning.

\section{Conditional VAIM (C-VAIM)}
\label{sec:C-VAIM}
It is worth noting that while a multitude of CFF solutions satisfy the scattering formula in Eq.~(\ref{eq:inverse-prob}), only one of them corresponds to the physical CFFs for a given set of kinematic variables. Importantly, although the four complex CFFs are derived from independent GPDs, a consistent evolution of the CFFs under variation of the kinematic variables is expected. Yet, for a VAIM model designed for a single set of kinematic parameters, the implicit correlations among CFFs at different kinematics cannot be captured. To incorporate this consistency condition, here we present a conditional VAIM (C-VAIM) model to be trained by CFF datasets from different kinematics, which is designed to derive CFFs satisfying the global kinematics constraints.

\begin{figure}[]
\includegraphics[width=0.99\columnwidth]{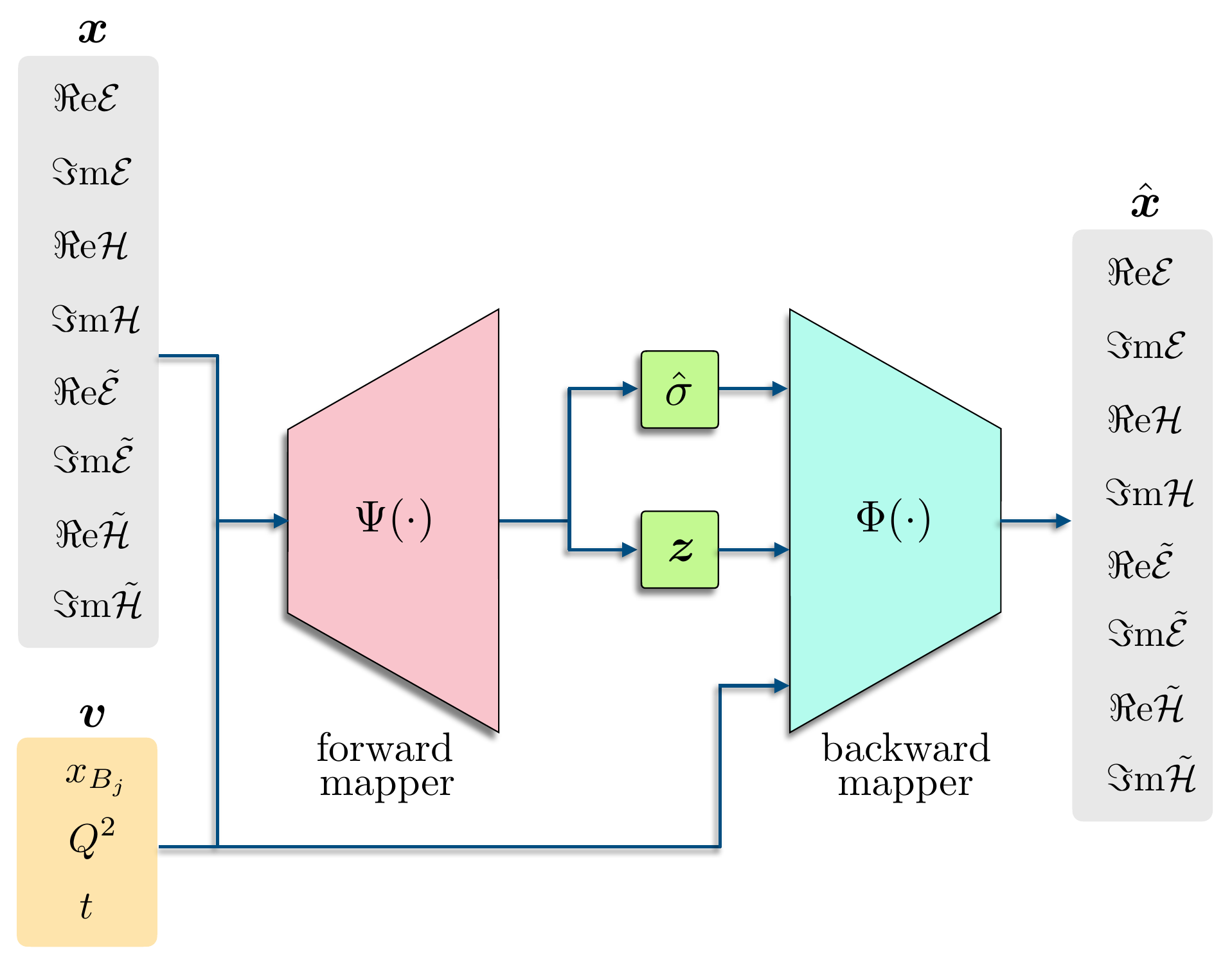}
\caption{\label{fig:c-vaim-scheme} Schematic diagram of the C-VAIM architecture for solving the inverse problem of CFF extraction. The basic structure is similar to a VAIM, with the important modification that both the forward and backward mapper neural networks are conditioned by the kinematic parameters $\bm v = (x_{Bj}, Q^2, t)$.}
\end{figure}

\begin{figure*}[]
\includegraphics[width=1.99\columnwidth]{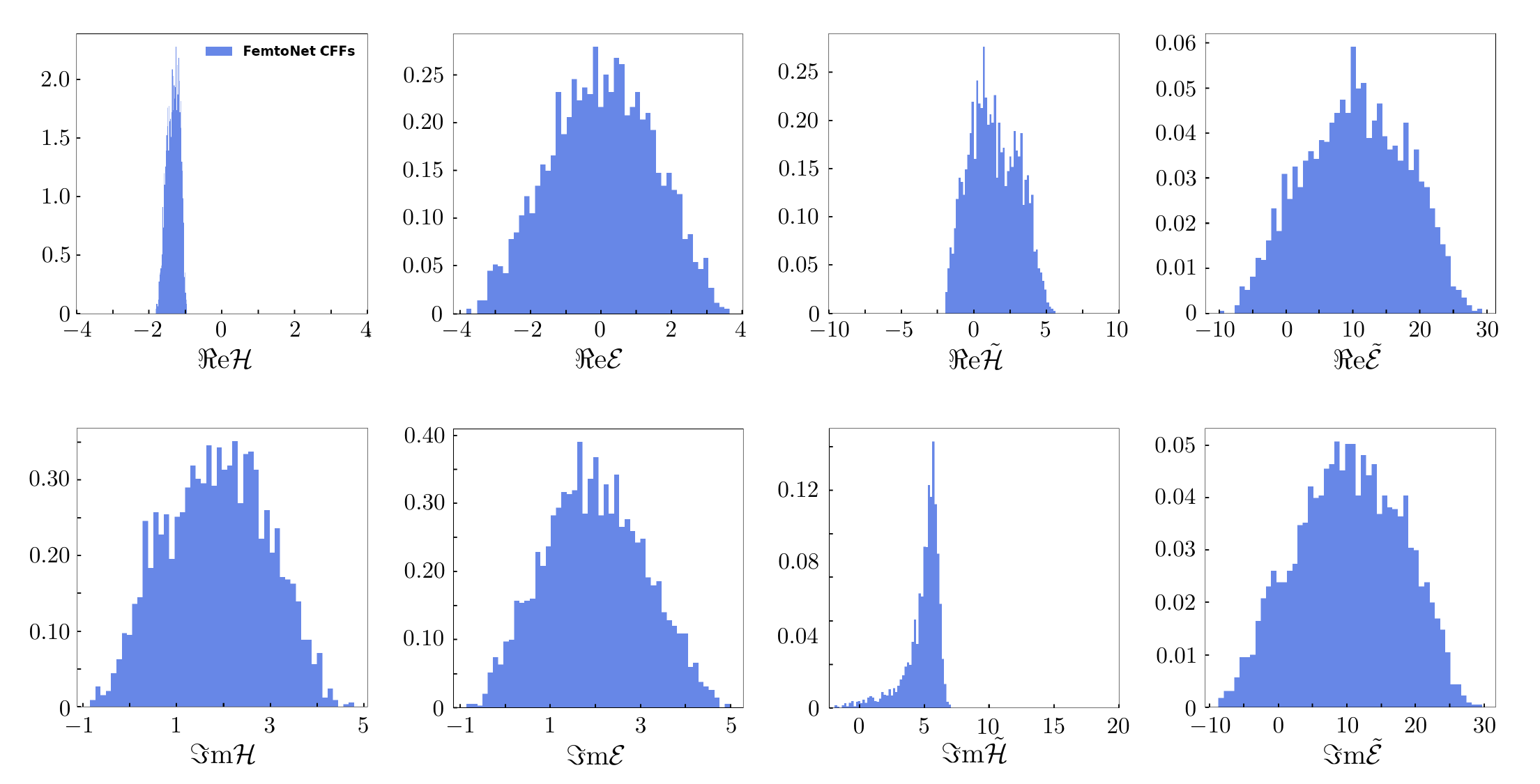}
\caption{\label{fig:hist2} Histogram of CFFs from sampling based on C-VAIM. All results are shown for a specific kinematics for the unpolarized cross section at $Q^2 = 1.82$~GeV$^2$, $t = -0.172$~GeV$^2$, $x_{Bj} = 0.343$, and $E_b = 5.75$~GeV.}
\end{figure*}

\begin{figure}[]
\includegraphics[width=0.99\columnwidth]{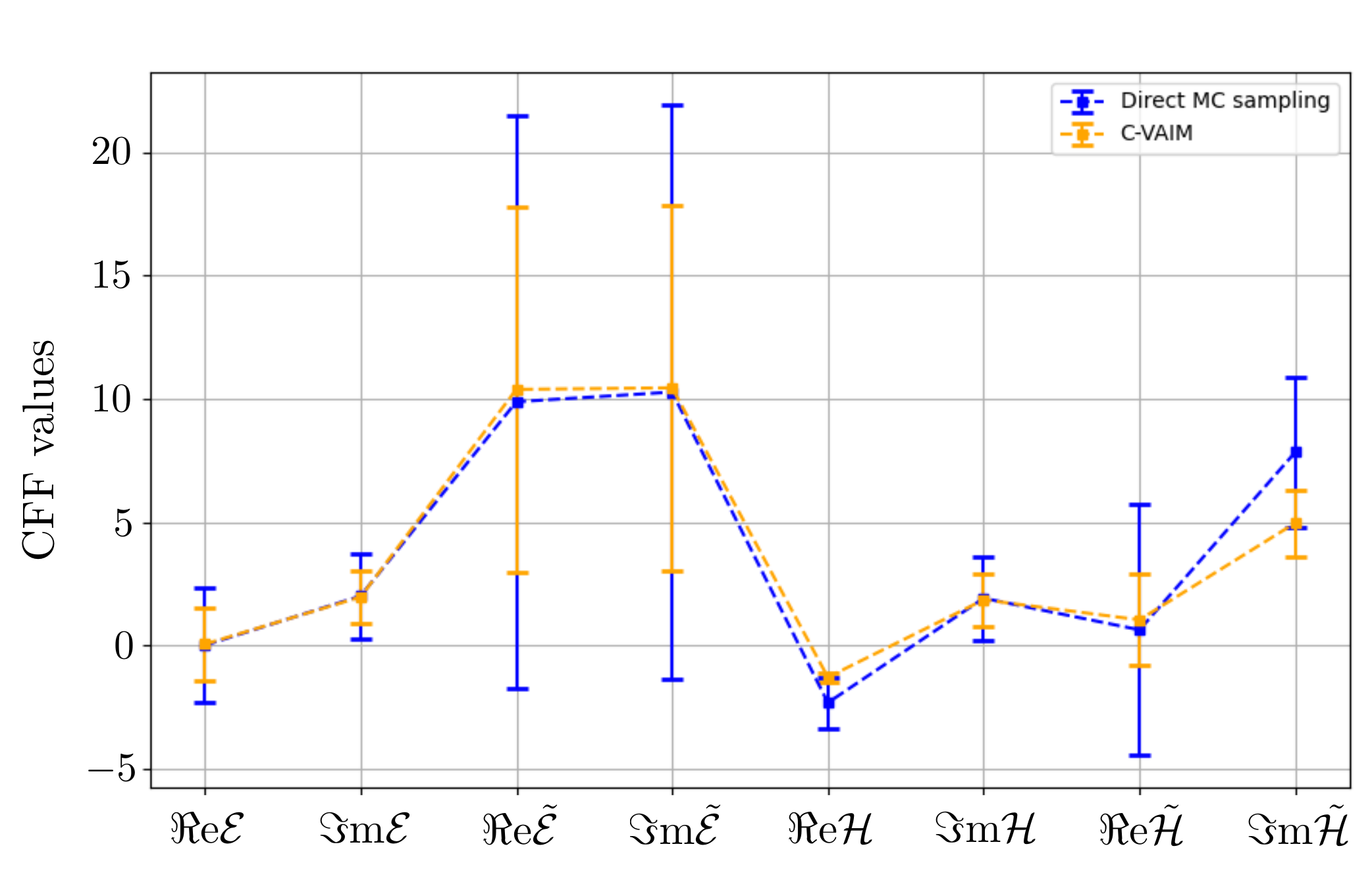}
\caption{\label{fig:cmp2} Comparison of CFF extracted from C-VAIM and direct Monte Carlo sampling.}
\end{figure}

\begin{table}[t]
\begin{center}
    \begin{tabular}{ |c | c| c | c | }   
    \hline
     Bin & $x_{Bj}$ & $t$ (GeV$^{2}$)& $Q^2$ (GeV$^{2}$)\\
    \hline 
    \hline
     1&  0.343& $-0.172$ & 1.820 \\
    \hline 
     2&  0.368& $-0.232$ & 1.933\\
    \hline
    3 & 0.375 & $-0.278$ & 1.964\\
    \hline
    4 &  0.379 & $-0.323$ & 1.986 \\
    \hline
    5 & 0.381 & $-0.371$ & 1.999\\ 
    \hline
    \end{tabular}
    \caption{Kinematics values used to generate datasets for the training of C-VAIM. }
    \label{tab:c-vaim-kinematics}
\end{center}
\end{table}

A schematic of the C-VAIM architecture is shown in FIG.~\ref{fig:c-vaim-scheme}. Similarly to the VAIM structure discussed in Sec.~\ref{sec:vaim-basics}, the model consists of two fully connected neural networks which implement the forward and backward mappers, respectively. The main difference is that the kinematic variables $\bm v = (x_{Bj}, Q^2, t)$ are input to both the forward and backward mappers. This structure is also similar to the standard C-VAE where the encoder and decoder are both conditioned on additional parameters~\cite{NIPS2015_8d55a249}. Essentially, the forward mapper is trained to approximate the true posterior distribution $p( \bm z \, | \, \bm x, \bm v, \sigma )$ by a tractable one  $q( \bm z \, | \, \bm x, \bm v, \sigma )$, while the backward mapper learns to approximate the likelihood distribution $p( \bm x, \bm v, \sigma \, | \, \bm z )$. Again, using the variational inference method~\cite{vae}, the optimization of C-VAIM is translated into the minimization of the following loss function
 \begin{eqnarray}
	\label{eq:loss_func2}
	& & L = \norm{\sigma - \hat{\sigma}}_2^2 + \norm{\bm x - \hat{\bm x}}_2^2 \\
	& & \qquad + {\rm KL}\bigl(q(\bm z \, | \, \bm x, \bm v, \sigma) \, || \, p(\bm z \, | \, \bm v) \bigr). \nonumber
\end{eqnarray}
Here the first term represents the prediction error of the forward mapper, the second term is the likelihood error of reconstruction, and $p(\bm z \, | \, \bm v)$ is a true prior distribution conditioned on the kinematics $\bm v$. As in VAIM, an easy-to-generate distribution, such as normal or uniform distribution, is used for this prior distribution function. 
By training the C-VAIM using datasets from selected sets of kinematic variables, the backward mapper is expected to predict CFFs at a continuous range of kinematics through transfer learning. Practically, for a given cross section and kinematic variables $\bm v$, a latent variable $\bm z$ is sampled from $p(\bm z\, | \, \bm v)$; these are then fed into the backward mapper neural network which produces CFFs at the output. 

To demonstrate the reduction of prediction uncertainty due to conditional learning, we trained a C-VAIM using CFFs sampled from exactly the same prior regions, summarized in Table~\ref{tab:prior-regions}, as those used in the training of VAIM. CFF and cross section datasets $\left\{ \bm x^{(m)}, \sigma^{(m)} \right\}$ at 5 different sets of kinematic parameters, shown in Table~\ref{tab:c-vaim-kinematics}, are used to train the C-VAIM. The hyperparameters of the neural networks in C-VAIM such as the number of neurons, number of layers, and the learning rate, are selected using Keras tuner~\cite{omalley2019kerastuner}. The optimized forward and backward mappers are both composed of three fully-connected hidden layers each with 1024 neurons activated by a leaky ReLU function. The network is regularized by an L2-norm penalty and a dropout rate of 0.2 to prevent overfitting.

The successfully trained C-VAIM is used to extract CFFs from unpolarized DVCS cross section at  kinematics $Q^2 = 1.82$~GeV$^2$, $t = -0.172$~GeV$^2$, $x_{Bj} = 0.343$, and $E_b = 5.75$ GeV, which are exactly the same as those used in the  VAIM and MCMC study in Sec.~\ref{sec:benchmark-vaim}. Histograms of the 8 sampled values are shown in FIG.~\ref{fig:hist2} with the range of CFFs set to their respective regions prior. The results indicate that the predicted CFFs are well constrained compared with the range used to sample the CFF training datasets. More importantly, in stark contrast to the case of VAIM and MCMC methods where nearly uniform distributions are obtained for several of the extracted CFFs (see FIG.~\ref{fig:hist1}), the predictions of the C-VAIM show nontrivial structures with clear peaks indicating the most likely CFF values. This is particularly true for the three CFFs $\Re e \mathcal{H}$, $\Re e \widetilde{\mathcal{H}}$ and $\Im m \widetilde{\mathcal{H}}$. As discussed in the previous section, some nontrivial features can already been seen even in the naive VAIM and MCMC predictions. These features due to the kinematic constraints are then dramatically enhanced by C-VAIM.  The significantly reduced uncertainties are also illustrated in FIG.~\ref{fig:cmp2} which compares the central values and standard deviations of CFFs obtained from C-VAIM and the MCMC methods. Overall, the uncertainties from C-VAIM is roughly 2/3 of those obtained from naive VAIM or MCMC.

It is worth noting that the idea of C-VAE was originally developed to facilitate a more controlled data generation process. Take the example of using VAE as a generative model for producing handwritten images from unicode characters. In a standard VAE,  there is no control over what characters will be generated by the decoder even for a well-trained model. This additional information can be supplemented by conditioning both encoder and decoder on extra input parameters. Interestingly, in the context of C-VAIM, the conditioning introduces further constraints on the predicted CFF values. Intuitively, this can be attributed to the fact that the forward and backward mappers are trained to simultaneously satisfy constrains from different kinematic variables. And since the neural networks depend on these conditioning variables $\bm v$ continuously, or in a ``differentiable" manner, the overall predictions are also expected to change continuously upon varying the kinematic variables. In other words, the VAIM learns the CFF manifold constrained by global kinematics.

\begin{figure}[]
\includegraphics[width=0.93\columnwidth]{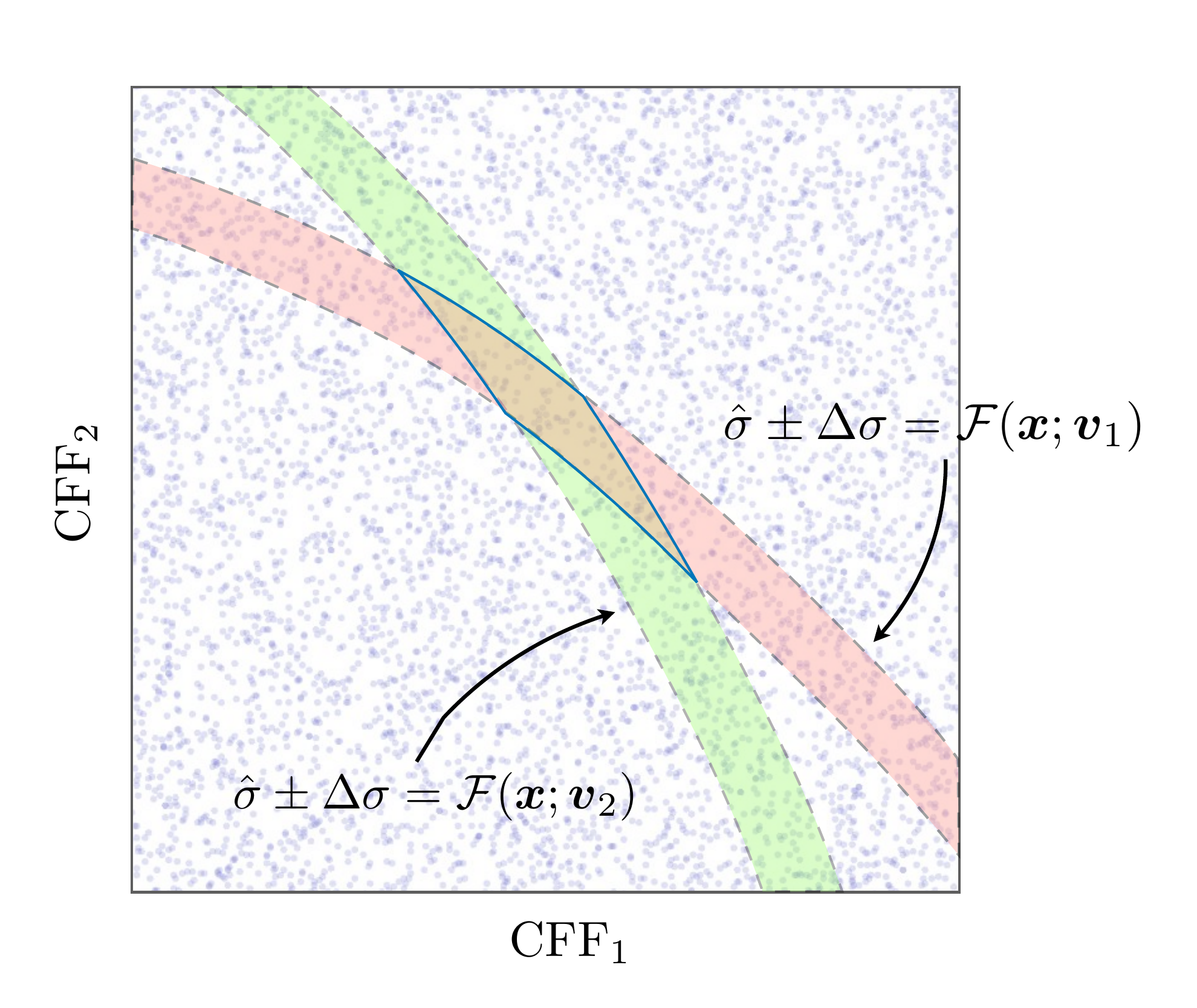}
\caption{\label{fig:conditioning} Schematic showing the constraints on CFF extraction due to conditional learning.}
\end{figure}

To illustrate this point, we again use the hypothetical $D=2$ CFF space as an example. Consider the two curves shown in FIG.~\ref{fig:conditioning} corresponding to the cross section $\hat{\sigma}$ constraint in Eq.~(\ref{eq:inverse-prob}) for two sets of kinematics $\bm v_1$ and $\bm v_2$. Introducing a small range $\Delta \sigma$ for the cross section extends the two curves into two strips of a finite width. Assuming a small difference between the two sets of kinematic variables, i.e. $\bm v_2 = \bm v_1 + \Delta \bm v$. As discussed above, the physical CFFs, computed from the corresponding GPDs, are uniquely determined by kinematic variables. The small difference between $\bm v_1$ and $\bm v_2$ thus means that the corresponding physical CFFs $\bm x_1$ and $\bm x_2$ are also in close proximity to each other. So does the resultant respective cross sections $\sigma_1$ and $\sigma_2$. These considerations point to a higher probability for the sampled CFFs to lie in the overlap region in FIG.~\ref{fig:conditioning}. Instead of uniformly distributed on the respective curves or strips, a more constrained distribution is obtained from C-VAIM prediction. Of course, these additional constraints depend crucially on the overlap geometry of the constrained manifolds. Back to the CFF extraction from real DVCS, comparison of histograms in FIG.~\ref{fig:hist1} and \ref{fig:hist2} such further constraints from C-VAIM convert the nearly uniform distribution in the prior domain into a broad-peaked distribution in the same domain for most of the CFFs. On the other hand, a highly structured distribution with a single sharp peak are observed for the three CFFs $\Re e \mathcal{H}$, $\Re e \widetilde{\mathcal{H}}$ and $\Im m \widetilde{\mathcal{H}}$.



\section{Conclusion and outlook}
\label{sec:conclusions}
Understanding the unpolarized cross section for DVCS -- the archetype process believed to be sensitive to GPDs in a factorized QCD picture -- is vital for setting up criteria for the study of any other DVES observable including different beam/target polarization configurations, and processes with additional particles other  than one photon in the final state. The extraction of CFFs from the unpolarized cross section has been the subject of many QCD analyses for the past two decades (see  Ref.~\cite{Kumericki:2016ehc} for a review of earlier attempts, and Ref.\cite{JeffersonLabHallA:2022pnx} and references therein) .

{In this paper, we first demonstrate that VAIM is consistent with MCMC in extracting multiple CFF solutions for a given set of kinematics. We then introduce C-VAIM, which generates CFFs constrained by the overall kinematics. C-VAIM effectively captures the correlations among CFFs with respect to different kinematic values, a task whihc is challenging for MCMC. Consequently, C-VAIM produces more constrained solutions compared to MCMC or VAIM under a single kinematics constraint.}



This is a necessary first step in the pipeline of our extended analysis that will go from  
the present cross section analysis and modelling/generalization including CFF extraction from observables 
solving the inverse problems with understandable errors to understand both physical integrated quantities, moments of EMT given by  angular momentum, pressure distributions, and spatial densities to the extraction of GPDs with their parameters.

\acknowledgements
This work was completed by the EXCLAIM collaboration under the DOE grant DE-SC0024644. The authors also thank the support of the SURA Center for Nuclear Femtography.

\appendix

\bibliography{ref}

\end{document}